\documentclass[aps,prl,twocolumn,groupedaddress,showpacs]{revtex4}
\usepackage{graphicx}
\newcommand{\gma}{$\rm{Ga_{1-x}Mn_{x}As}$}

\newcommand{\tc}{$T_{\rm{C}}$}

\begin{document}
\title{Measurements of Nanoscale Domain Wall Flexing in a Ferromagnetic Thin Film}
\author{A. L. Balk,$^1$ M. E. Nowakowski,$^2$ M. J. Wilson,$^1$ D. W. Rench,$^1$ P. Schiffer,$^1$ D. D. Awschalom,$^2$ N. Samarth$^{1}$}
\email{nsamarth@psu.edu}
\affiliation
{$^{1}$Department of Physics, The Pennsylvania State University, University Park, Pennsylvania 16802, USA \\
$^{2}$Center for Spintronics and Quantum Computation, University of California-Santa Barbara, California 93106, USA 
}
\date{\today}

\begin{abstract}

We use the high spatial sensitivity of the anomalous Hall effect in the ferromagnetic semiconductor \gma, combined with the magneto-optical Kerr effect, to probe the nanoscale elastic flexing behavior of a single magnetic domain wall in a ferromagnetic thin film. Our technique allows position sensitive characterization of the pinning site density, which we estimate to be $\sim$10$^{14}$cm$^{-3}$. Analysis of single site depinning events and their temperature dependence yields estimates of pinning site forces (10 pN range) as well as the thermal deactivation energy.  Finally, our data hints at a much higher intrinsic domain wall mobility for flexing than previously observed in optically-probed $\mu$m scale measurements.

\end{abstract}


\maketitle
Understanding the fundamental behavior of magnetic domain walls (DWs) in ferromagnets \cite{Tatara:2004dq,Beach:2005pd,Marrows:2005ve,Yamanouchi:2007bs} continues to attract significant attention  because of potential spintronic applications in memory and logic \cite{Parkin:2008qa,Allwood:2005bh}.  As first postulated by Ne\'el \cite{Neel1946}, DWs are not rigid: they follow the physical laws of an elastic interface that interacts with a disordered potential comprised of spatially localized pinning sites. The interplay between elasticity and pinning directly influences macroscopic properties such as coercive field and hysteresis shape \cite{Jiles:1986bh,Jatau:1995ai}.  For low applied magnetic fields, elasticity and pinning manifest as non-repeatable thermally activated DW creep when the interface stochastically jumps from pinning site to pinning site \cite{Lemerle:1998uq}. At higher magnetic fields, DWs propagate in the ``flow'' regime, although their mobility is still limited by viscous drag arising from pinning sites \cite{Dourlat:2008ve,Tang:2004uq}.  Significant efforts have been directed towards measuring and controlling the behavior of single DWs via diverse techniques such as magneto-optical Kerr effect (MOKE) imaging \cite{Beach:2005pd,Marrows:2005ve,Yamanouchi:2007bs,Lemerle:1998uq}, magneto-resistance \cite{Tang:2004uq}, scanning Hall magnetometry \cite{K.-S.-Novoselov:2002fk} and x-ray microscopy \cite{HyunKim:2005}. However, none of these have provided a method to directly measure the elastic flexing of a single DW, a phenomenon long predicted by Ne\'el \cite{Neel1946} but not yet directly observed, and of fundamental importance to understanding DW propagation and mobility. 

Here, we exploit the anomalous Hall effect (AHE) to measure the position of a magnetic DW to nanometer precision. This allows us to directly probe the reversible nanoscale flexing of a single DW. In this flexing regime, we find that DWs exhibit different kinematics from the better studied regimes of creep and flow. A key finding is the observation of a large intrinsic DW mobility that far exceeds the values deduced from earlier studies at $\mu$m length scales. Finally, with a simple geometric model to describe the flexing behavior of the DW, we estimate the pinning site density, strength, and energy. The methodology demonstrated in this manuscript is generic and can be readily extended to ferromagnetic materials other than the specific one used in our study. 

Our measurements are carried out on micro-fabricated devices patterned from a 25 nm thick epitaxial layer of \gma~under in-plane tensile strain, which creates samples with perpendicular magnetic anisotropy. The choice of the ferromagnet is principally driven by the large AHE in this material which makes the measurements particularly convenient. The \gma~samples are grown via low temperature molecular beam epitaxy on top of a relaxed buffer layer of In$_{\rm{x}}$Ga$_{\rm{1-x}}$As, itself grown on a semi-insulating substrate of (001) GaAs (Fig. 1a. inset). The growth conditions are similar to those described elsewhere \cite{Xiang:2005yk}. The samples are then annealed at $190^{\circ}$C for 120 hours in air, which increases the sample Curie temperature (\tc) (Fig. 1 (a)), while simultaneously increasing the magnetic homogeneity of the sample \cite{Dourlat:2007vn}. Finally, devices are patterned into Hall bar structures using a standard wet etch process. The voltage probes are patterned from the sample material itself using electron beam lithography. Ohmic electrical contacts are made using indium and the anomalous Hall effect is measured using standard phase sensitive ac techniques in a He flow cryostat with an external magnetic field. Additionally, simultaneous measurements employing a video-rate MOKE imaging system are used to calibrate DW positions with AHE measurements. We discuss detailed measurements on two devices patterned from the same sample: a 10$\mu$m and a 20$\mu$m wide Hall bar (Fig.1(b)).  

\begin{figure}[]
\includegraphics[width=70mm]{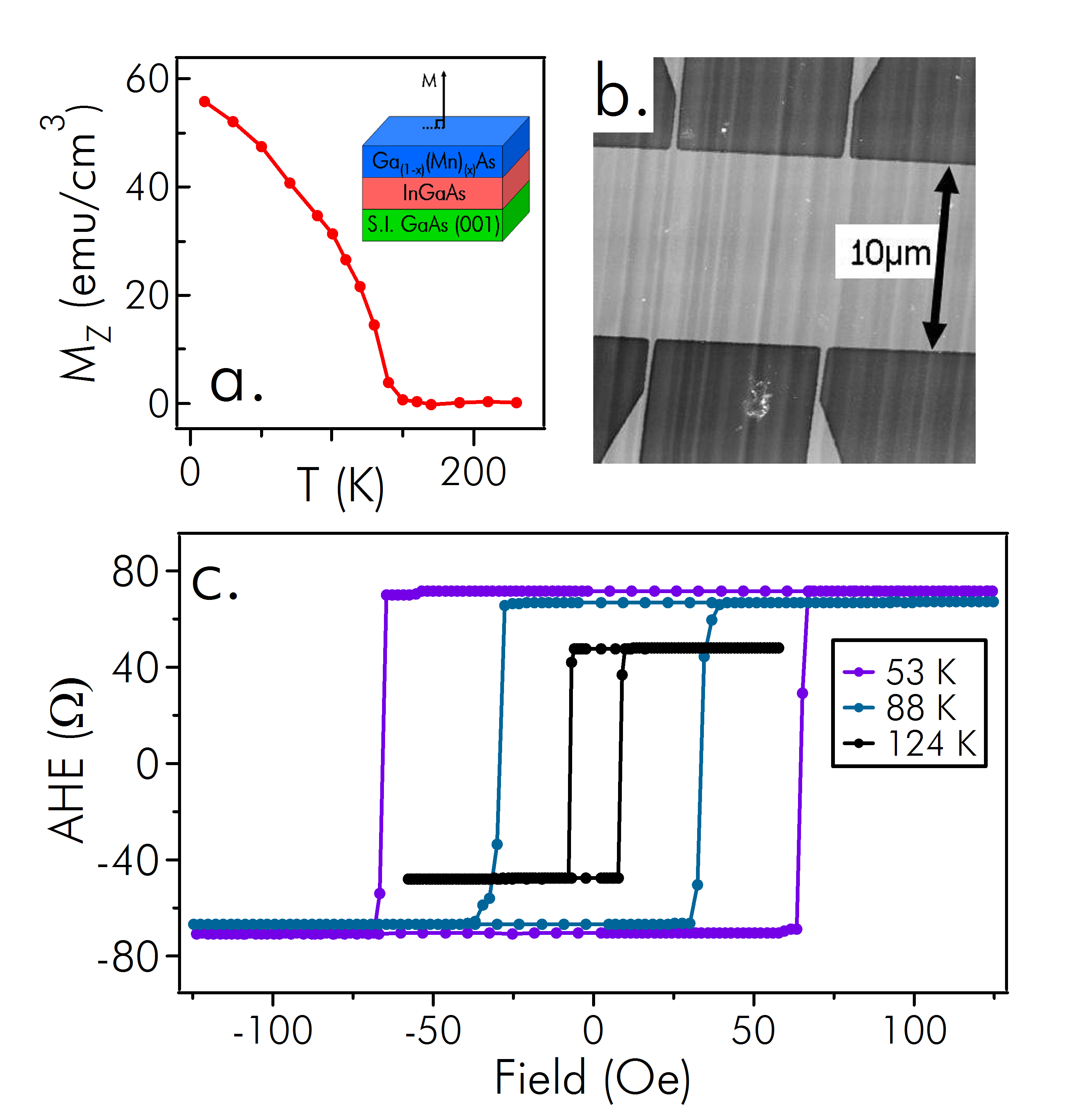}
\caption{a. Magnetization versus temperature of an unpatterned sample measured using a SQUID magnetometer, showing \tc$=140$ K. Inset: Schematic of the sample structure.  b. Atomic force microscope scan of the 10 $\mu$m device.  c. Hysteresis loops measured with AHE at $T =$ 53 K, 88 K, and 124 K.}
\end{figure}

The AHE can be used as a sensitive local probe of the sample magnetization (for example, Fig. 1(c)). In ferromagnetic thin film samples with perpendicular magnetic anisotropy, \cite{Cheng:2005rr,Xiang:2007pf} the AHE can be correlated with the position of a DW. If a single vertical Bloch DW is present in a Hall bar of width $\emph{a}$, the magnitude of the anomalous Hall resistance at a distance $x$ from the DW is given by:
\begin{equation}
\frac{[AHE]} {[AHE]_{\rm{SAT}}}={\bigg(}1-\frac{8} {\pi^2}\sum_{n=odd}^{\infty}{\frac{e^{\frac{-\pi{n}\mid{x}\mid} {a}}} {n^2}} 			{\bigg)}
\end{equation}
Due to the large magnitude of the AHE in \gma, the average measurement sensitivity is on the order of 10 $\Omega$/$\mu \rm{m}$, over the roughly 20$\mu \rm{m}$ measurement area,  translating into a maximum measurement standard deviation of $\lesssim 2$nm for measurements on the order of seconds. 

To assess the reliability of correlating the measured AHE with the DW position in our material, we carried out simultaneous MOKE and AHE measurements.  First, we used MOKE images to verify that the device has a simple magnetic domain structure with a single DW under typical conditions used for the AHE measurement to be discussed later. We then compared the AHE with the DW position as measured by MOKE; although this measurement is carried out over a large range of DW position (see supplementary material), we note that the theoretical correlation given by Eq. 1 is only valid for DWs within a limited range ($\sim$20 $\mu$m) of the contacts. Figure 2(a) plots the theoretical expectation for two different Hall bar widths (dashed and solid lines) and compares this expectation with MOKE data taken for two different DW locations (insets, Fig. 1a) on the 10 $\mu$m device.  A few comments are warranted about the very conservative uncertainties depicted by the error bars in this figure. The theoretical expression in Eq. 1 assumes a single DW and magnetization reversal occurring solely due to DW motion, not nucleation.  These nucleation events are visible in the hysteresis loop as shoulders before and after the switching event (Fig. 1c), which become more prominent with the imaging light used for MOKE imaging (see supplementary materials). This leads to some ambiguity in determining the saturation values of the AHE if there is domain nucleation in the measurement area.  This effect, coupled with increased measurement noise due to the light, results in the conservative placement of error bars in Fig. 2a.  

\begin{figure}[]
\includegraphics[width=70mm]{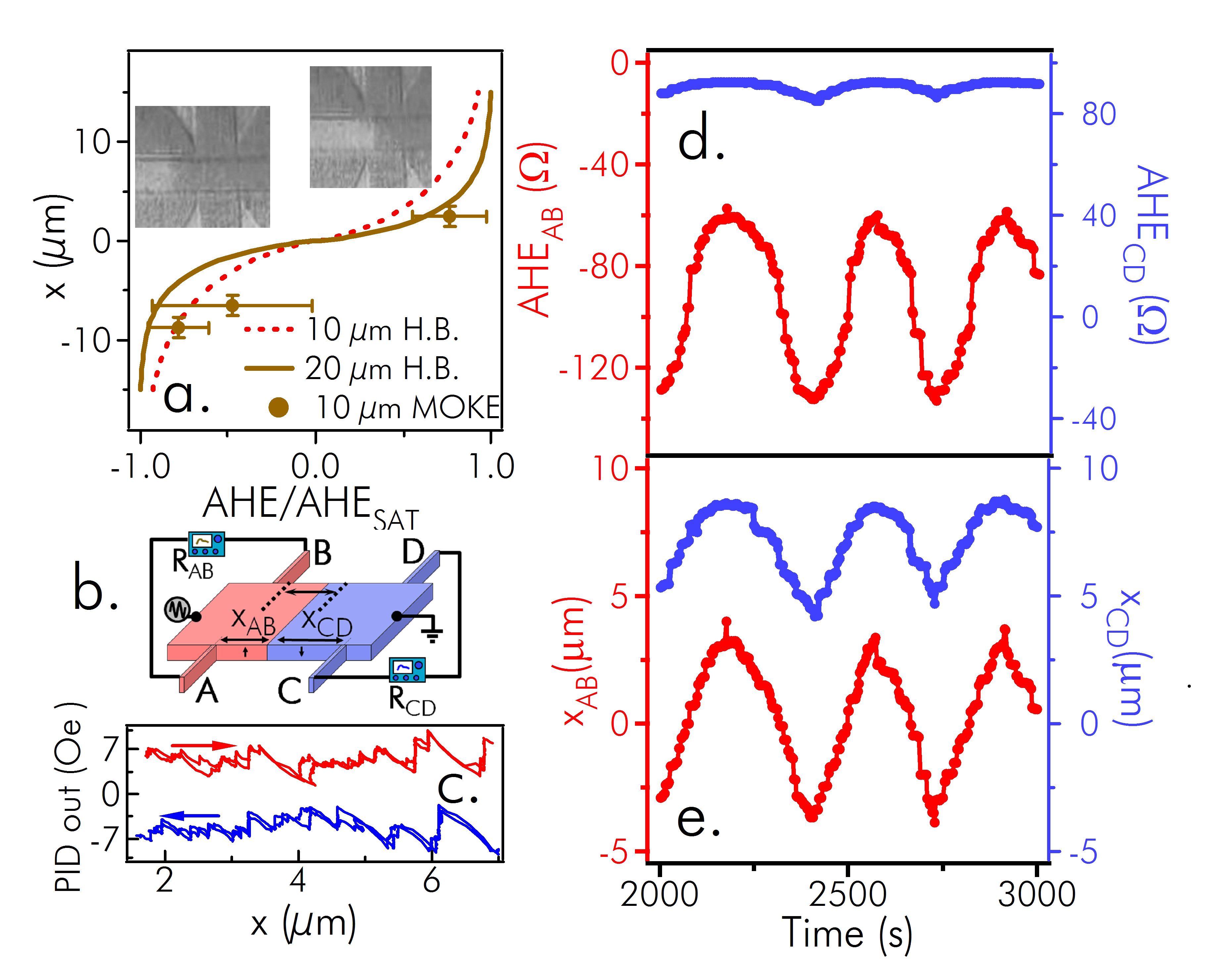}
\caption{a. Measurements of the DW position in the 10 $\mu$m channel device as determined by MOKE imaging plotted as a function of simultaneously measured AHE.   Theoretical expectations (dashed and solid lines) are also plotted as a reference.  Insets: MOKE images used to obtain data points.  b. Schematic of measurement setup.  c.  Magnet output required to scan the DW at constant velocity plotted against electrically measured position, showing DW pinning sites (sharp peaks).  d.  Measurements of the AHE vs. time from two sets of Hall probes while the DW is being scanned. DW. e. Data in \textit{d} converted to position, showing correlation between the two measured positions.}
\end{figure}

We first take advantage of the correlation between AHE and DW position to study the basic DW pinning behavior of the sample.  We use the AHE and Eq. 1 to measure the DW velocity from one set of Hall probes (probes $AB$, Fig. 2b) and raster it between set positions (dashed lines in Fig. 2b) at a constant speed of 150nm/sec with a proportional-integral-derivative (PID) loop and a trimming electromagnet.  By plotting the output of the PID loop as a function of DW position (Fig. 2(c)), we obtain estimates of the spatially resolved DW depinning fields (sharp peaks in 2(c), 2 Oe $\lesssim H_{\rm{depin}}\lesssim$ 10 Oe).  For our current experimental setup, the measurement is not fast enough to smoothly control the DW velocity when it depins, so any jumps in potential the DW encounters after depinning will not be manifest in the data, and thus we cannot take the distribution of peaks in this data as representing the true density of pinning sites, but a lower limit.

This procedure further verifies the correlation between AHE and DW position. While scanning the DW, we simultaneously monitor the AHE from two sets of Hall probes, the control set R$_{\rm{AB}}$ and a reference set (R$_{\rm{{CD}}}$).   In this measurement, the DW is maintained closer to probes AB.  As expected, the resistance trace from probes CD shows both a lower sensitivity to DW position and an overall offset, as would be expected from a single DW picture.  When converted from resistance to DW position (Fig. 2e), the data traces are offset from each other by $\approx$7 $\mu$m, which is close to the lithographically defined 10 $\mu$m probe separation.  The difference in these values arises from the difficulty in accurately determining AHE$_{\rm{SAT}}$ values; this error manifests as an offset in the measured position which is most significant at large separations between the Hall probes and DW. For the additional data presented in this study, we ensure that the DW is between 0.3 $\mu$m and 8 $\mu$m from the Hall probes.  The upper limit minimizes the effect the incorrectly determined AHE$_{\rm{SAT}}$ values have upon the measurement, and the lower limit is used to eliminate any interaction between the DW and the Hall probes themselves. 

\begin{figure}[]
\begin{center}
\includegraphics[width=70mm]{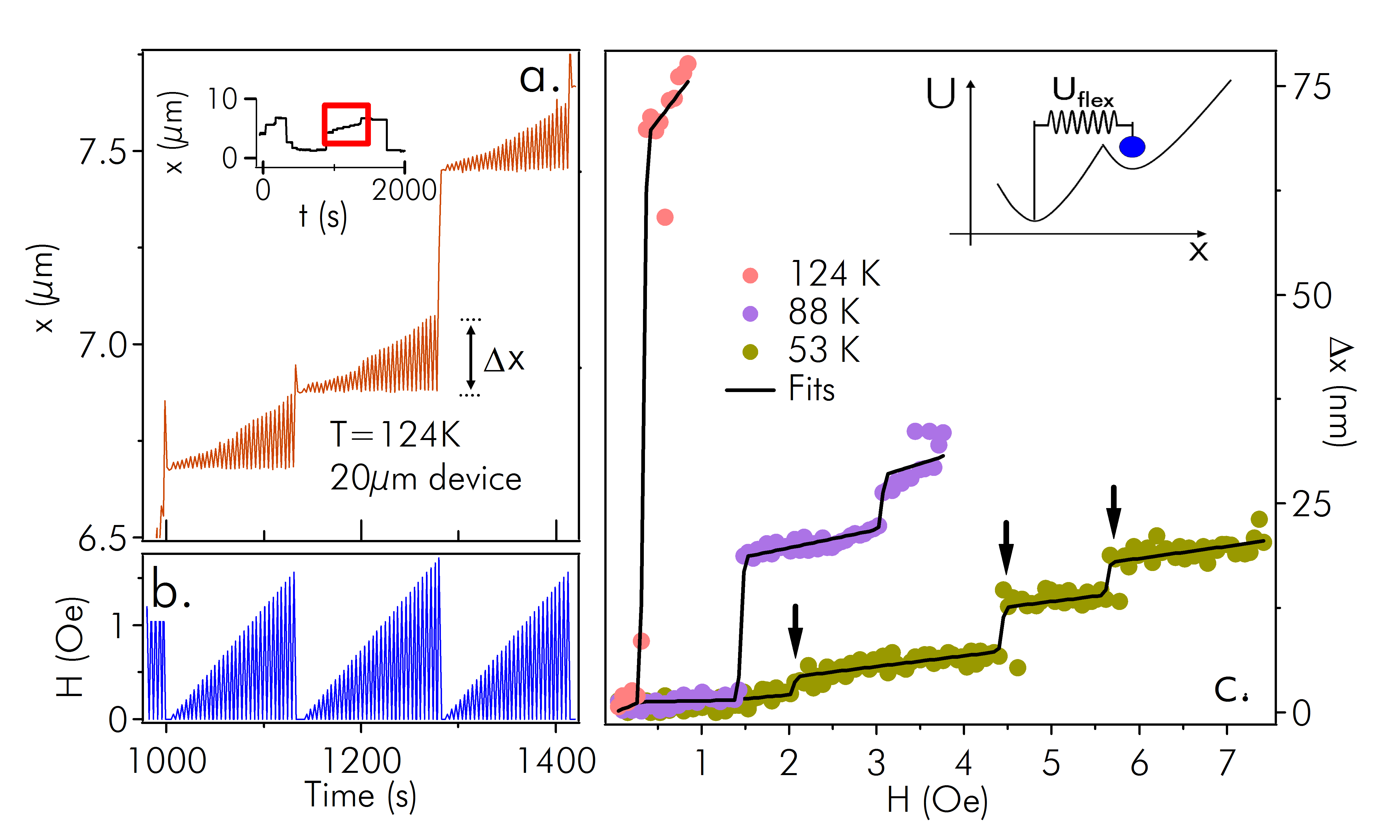}
\end{center}
\caption{a. Electrically measured DW position vs time, showing three remagnetization measurements. Inset: Position of DW over a longer timescale, sweeping across the measurement area, with three measurements shown in this panel boxed.  b. Field applied to obtain trace in \textit{a}.  c. Change in DW position $\Delta$x between the field applied and field zero states, demonstrating adiabatic, reversible remagnetization (linear areas) and sharp depinning events (indicated by arrows in the $T = 53$ K trace). Inset: Cartoon of conservative potential required to obtain a sudden change in DW position.  For the DW to return to the original position, there must be an additional restorative potential U$_{\rm{flex}}$.}
\end{figure}

To measure DW flexing, we search for adiabatic, low field DW motion which reverses upon removal of the field.  After cooling the sample below \tc~ and obtaining temperature dependent values of AHE$_{\rm{SAT}}$ (used to scale Eq. 1), we generate a DW by slowly ramping a magnetic field which is set to zero when the AHE starts to move from its saturation value.  We then use the PID loop to verify a single DW state and establish a $\sim$ 6$\mu$m length along the device with no strong pinning sites.  Next, we position the DW within this area, apply a small field (0 Oe $\lesssim H \lesssim$10 Oe) to the Hall bar, and measure the new position with the field applied after a time delay $\Delta t$.  If the DW returns to within a preset limit (40 nm) of its original position after the field is removed, we repeat the procedure with a higher field. (Fig. 3(b))  The field is ramped in this fashion until the DW fails to return to its original position, at which point the experiment is repeated starting at the new position. The control program also switches the direction of the applied field when the DW moves outside the measurement area (2 $\mu$m$< x < $ 8 $\mu$m) in either direction (Fig. 3(a) inset).  For small fields, we observe repeatable, adiabatic remagnetization (Fig. 3(a)). When subtracted from the original positions, the remagnetization distance $x$ is linear with respect to applied field for low temperatures ($T \lesssim 90$ K).  At higher temperatures (90 K $\lesssim T < 120$ K) and fields, linear regions in the traces are separated by non-adiabatic jumps over tens of nanometers (arrows, Fig. 3c).  This data indicates that we are observing a different regime of DW motion than thermally activated creep, which would show up as an exponential velocity dependence upon applied field \cite{Dourlat:2008ve}.

Reversible DW motion can be explained by two processes: DW flexing (small, strong pinning sites) and conservative DW motion within sites (large, weak pinning sites). The dominant behavior depends on the DW surface tension $\sigma = \sqrt{A K_u}$ ($A$ and $K_u$ are the spin stiffness and uniaxial magnetic anisotropy constants, respectively), pinning site density $\lambda$, and pinning site anisotropy.  Flexing will dominate for a stretchy wall (low $\sigma$), low $\lambda$, and high pinning site anisotropy.  We choose to use a DW flexing model based upon four observations.  

First, the PID output in Fig. 2(c) does not repeat exactly upon different trips across the device, suggesting the DW is bending to visit different pinning sites upon each trip for this feedback measurement.  Second, the large spread in depinning fields presented in Fig. 2(c) suggests the DW could be depinned from a portion of sites along its width while remaining pinned at others.  We interpret the sudden jumps (arrows, Fig. 3c) in the otherwise adiabatic data as such depinning events.  Furthermore, if these are depinning events, they cannot be due to conservative DW motion.  If this were the case, the restoring force acting on the DW would have to be less than zero at the discontinuity (i.e. push the DW farther) and thus the DW would not return to its original state, ending the measurement run (Fig. 3c, inset).  In fact, we see up to three discontinuities in the data traces, indicating an additional restorative potential U$_{\rm{flex}}$.  Finally, we note that there is a strong temperature dependence of  the maximum reversible domain wall displacement (Fig. 3c), which would unrealistically indicate a dependence of pinning site size on temperature if the dominant effect were motion within pinning sites. 

The data is analyzed using a simple geometrical picture that assumes a vertical Bloch DW with a semicircular profile in between pinning sites \cite{Jiles:1986bh}. This model reflects an energy balance between the Zeeman energy of the applied field and the elastic energy from the surface of the distended DW.  Assuming the bending distance is very small compared to the lateral width of the DW and the pinning site separation, the model produces the measured linear relationship between the applied field and the distance through which the DW bends.  The slope of the linear relationship $m$ is related to the sample magnetization $M_z$, the pinning site separation $y$, and the DW energy via the relation: ${y}={4}[\frac{{m}\sqrt{{A}{K_u}}}{M_z}]^{\frac{1}{2}}$. We obtain the uniaxial anisotropy constant from the magnetization $K_{u} = 11 {M_{S}}^{2}$ \cite{Wang:2005zs}.  The spin stiffness constant $A$ is determined using the temperature independence of the DW width $\delta = \pi\sqrt{\frac{A}{K_u}}\approx 15$ nm, a value we obtain from \cite{PhysRevB.64.241201}. To measure the DW elasticity $m$, we fit the data in Fig. 3(c) to linear areas separated by sigmoid functions which model the sudden depinning events.

To confirm that the data has no contribution from the onset of thermally activated creep (irreversible motion), we carry out a time-dependent measurement in which the duration of the applied field ($\Delta t$) is varied from 1 s to 15 s.  This data is presented in Fig. 4(a) for the 20 $\mu$m device at $T = 106$ K for four representative $\Delta t$ values.  If creep were a factor in these traces, there would be a direct correlation between the slopes $m$ of the linear areas of these traces and their corresponding $\Delta t$ values.  At this temperature, we observe creep only at $\Delta t > 12$ s, (Fig. 4a. inset).  For the additional data in this study we limit the value of $\Delta t \leq 12$ s, and to confirm creep is not a factor for higher temperatures, we verify the DW position before and after field applications (supplementary materials). 

These time dependent measurements are limited by the minimum response time of the lock-in amplifier, and thus the DW flexing mobility cannot be accurately measured with the current experimental setup.  However, noting the similarity in the values of $m$ between the $\Delta t = 1$ s and $\Delta t > 1$ s traces (Fig. 4a inset), we can place a lower limit on the mobility with the following argument.  The total time required to measure the equilibrium position of the DW after field application is the convolution of two factors: the intrinsic DW response time $\tau_{DW}$ and settling behavior of the measuring device.  Based upon the settings of the lock-in, we can calculate the theoretical measured response to an instantaneous DW displacement ($\tau_{\rm{DW}} = 0$).  This is represented as the blue curve in Figure 4b.  Any possible DW path will have to take place above the blue curve to avoid differentially affecting the $\Delta t = 1$ s and $\Delta t = 2$ s measurements (green and pink circles) which are within experimental error of each other.  The slowest response is determined as the line tangent to the blue curve (red line) which provides an upper $\tau_{\rm{DW}}$ and thus lower mobility limit of 40 nm/Oe.s.   This limit represents a factor of four increase over the same material in the literature for similar reduced temperature \cite{Dourlat:2008ve}.  In fact, since this flexing motion is not dissipative as creep, we expect the true mobility to be much higher as it will only be moderated by intrinsic factors. 

\begin{figure}[]
\includegraphics[width=70mm]{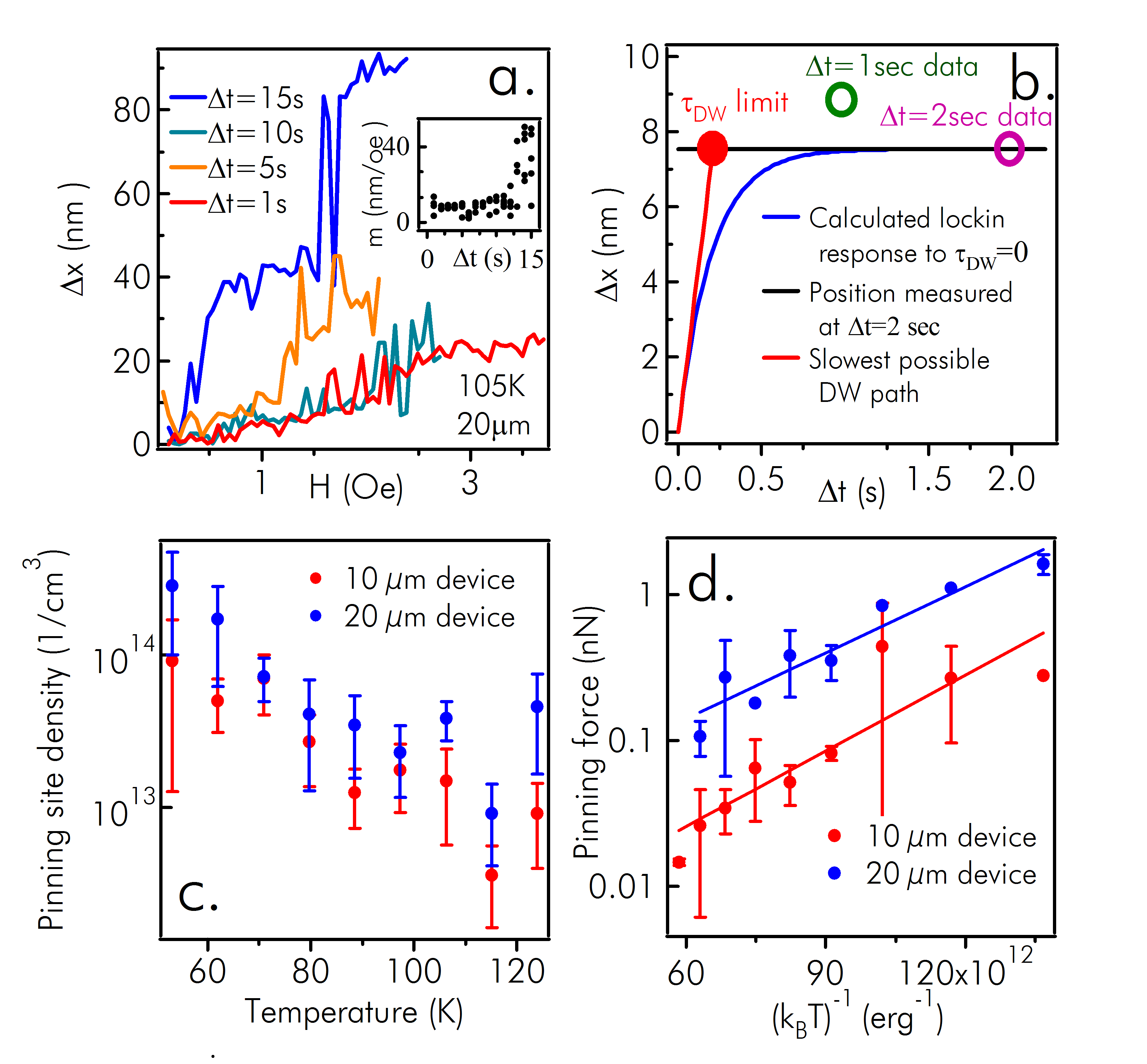}
\caption{a. $\Delta x$ measured for $\Delta t =$ 1, 5, 10 and 15 seconds at $T = 88$ K on the 20 $\mu$m device.  The similarity of the slopes at the low field limit rule out DW creep contributions to the measured DW flex distance.  Inset: Measured slopes plotted against $\Delta$t showing the onset of DW creep at $\Delta t >$12 s.  b. DW displacement for a field application of 1 Oe showing the time response of the measurement.  The measured displacements for $\Delta t = 1$ s and $\Delta t = 2$ s (open circles) are within experimental error of each other.  This implies the minimum time for the DW to reach equilibrium is represented by the tangent (red line) to the theoretical lock-in response (blue curve)   c. Pinning site density $\rho$ measured at a constant $\Delta t = 8$ s for both devices, showing a decrease in pinning site density as a function of increasing temperature.  d. Arrhenius plot of the pinning force calculated with Eq. 3, demonstrating the thermal deactivation energy for the pinning sites.}
\end{figure}

We then carry out high precision measurements at a constant $\Delta t= 8$ s for both devices over a range of temperatures.  At the lowest temperatures measured, we calculate a typical low temperature pinning site density to be in the 10$^{14} {\rm{cm}}^{-3}$ regime (Fig. 4c). We note that at low temperatures the pinning site separations are within errors of each other, demonstrating that in this regime of hall bar width the edges of the device do not greatly affect the DW behavior.  The density also shows a strong, consistent dependence upon temperature, suggesting that pinning sites become thermally deactivated at high temperatures. To confirm this, we use the field dependence of the depinning events (arrows in Fig. 3c) to measure the temperature dependent depinning force $f$, using a theoretical picture of strong DW pinning by sites much smaller than their separation\cite{Gaunt:1983fk}. This model yields a critical depinning field $H_{\rm{depin}} = [3 \rho {f^2}]/[8 \pi {M_z} \sqrt{A K_u}]$, where $\rho$ is the inverse of the pinning site separation.  When plotted against $(k_{\rm{B}} T)^{-1}$ (Fig 4d), the pinning force data demonstrates thermal deactivation with an associated activation energy of $21 \pm 2$ meV for the 20 $\mu$m Hall bar, and $25 \pm 5$ meV for the 10 $\mu$m Hall bar.  The difference in average pinning force between the two Hall bars is presumably due to a small difference in pinning site density between the two devices (Fig. 4c.).  

In summary, we have developed a novel technique to electrically measure the position of a magnetic DW to nanometer precision,  finding evidence for reversible elastic DW deformation which follows different kinematics from better studied regimes of DW motion.  A simple geometric model to describe the flexing behavior of the DW allows estimates of the pinning site density, strength, and energy. The observation of an enhanced DW mobility (which exceeds the current measurement capability of our instruments) has important implications for spintronic applications based upon DW manipulation.

This work was supported by ONR MURI under grant N0014-06-1-0428 and by NSF under grants DMR-0801406 and -0801388.






\end{document}